# WEATHER DATA ANALYSIS BASED ON TYPICAL WEATHER SEQUENCES. APPLICATION: ENERGY BUILDING SIMULATION.

Mathieu David, Laetitia Adelard, François Garde, Harry Boyer
Laboratoire de Génie Industriel, University of La Reunion,,
40 Avenue de Soweto, 97410 Saint-Pierre, France
Email : mathieu.david@univ-reunion.fr

## ABSTRACT

In building studies dealing about energy efficiency and comfort, simulation software need relevant weather files with optimal time steps. Few tools generate extreme and mean values of simultaneous hourly data including correlation between the climatic parameters.

This paper presents the C++ Runeole software based on typical weather sequences analysis. It runs an analysis process of a stochastic continuous multivariable phenomenon with frequencies properties applied to a climatic database.

The database analysis associates basic statistics, PCA (Principal Component Analysis) and automatic classifications. Different ways of applying these methods will be presented. All the results are stored in the Runeole internal database that allows an easy selection of weather sequences. The extreme sequences are used for system and building sizing and the mean sequences are used for the determination of the annual cooling loads as proposed by Audrier-Cros (Audrier-Cros, 1984).

This weather analysis was tested with the database of the French weather forecast utility Meteo France. Reunion Island experiences a lot of different micro-climates due to the high altitude (3069m), specific relief, and geographic situation (Tropic of Capricorn). Furthermore Reunion Island has the densest meteorological network in France and is an ideal place to validate the methodology with different climates.

To test the efficiency of such analysis, simulations using the resulting weather sequences were carried out with the building simulation software CODYRUN.

This analysis is the first step of a more global research concerning weather data generation. Future work will permit whole hourly typical meteorological year generation using neural networks.

## INTRODUCTION

The analysis of the behaviour of the energy systems such as the building requires the simultaneous knowledge of several climatic variables. The main climatic parameters are the temperature, the solar radiation, the humidity and wind conditions. These constraints must take into account the weather solicitations during the year and the day. Systems sizing and energy behaviour forecasting for a given site and for a given daily load profile requires a characterization of the random weather variation at this location. This study develops a methodology of "typical" weather sequences classification in order to elaborate a generator of synthetic weather sequences in view of systems sizing and energy forecasting. This method has been experimented for buildings simulations for different microclimates of Reunion Island (21°S, 55°E) (David, 2004) (Garde, 2005).

Most common weather generators for building simulators, Meteonorm (Remund, 2004), TRNSYS type 54 (Knight, 1991) (TRNSYS, 2000) and WGEN (Richardson, 1984) generate first the most important parameters. Generally this principal variable comes directly from radiation measurement. Meteonorm and WGEN (Remund, 2004) (Richardson, 1984) use the probabilistic transitions between thresholds of the variable to generate this variable from Markov Chains. And TRNSYS (Knight, 1991) use a stochastic model based on distribution laws for the long term average and a first order autoregressive model for the daily variation. The other parameters are generated by empirical or temporal models. An other way to generate synthetic series of days is to study the weather sequences or typical days.

Sacre (Sacre, 1984) uses Principal Component Analysis (PCA) to determinate the best factors to explain the daily evolution of the climatic parameters from data with shorter time step than a day. The 3 first fuzzy variables obtain with the PCA explain more than 90% of the standard deviation of one parameter during a day.

A study investigated in Trappes and Carpentras (France) (Boullier, 1984), based on the integration of PV system in house, used reduced variables to classify climatic data sets in typical weather days. This work presents the Ward's hierarchic cluster method with Euclidean distances to classify daily climatic data. At each level of the hierarchy one can identify a set of classes. Using this iterative method, the classification clusters the days of which the aggregation conduces to the lower inertia. The manual study of resulting levelized histogram permits to locate the homogeneity and the difference between groups of days.

From works previously quoted, Muselli reproduced the method of classification (Muselli, 2000) by using discriminant parameters and their derivative to characterize the direct daily radiation from hourly data. His analysis of the direct irradiance on inclined planes gives the probability of transition between the various classes of typical days.

We shall use the method of Sacre (Sacre, 1984) to express 3 fuzzy variables by PCA for each of the climatic parameters to study in the hourly step. This analysis allows the best possible description of a day with a minimum of parameters. The number of parameters has a great influence for the computation time.

From the results of the classification, we generate an artificial typical year. The aim of this paper is the validation of the selected classification method to produce pertinent entries for the generation process. In order to evaluate if the resultant typical year can replace a long term period, we compare the statistic parameters of the artificial annual weather file and a measured long term weather file. We lead the same comparison on a building simulation.

## INTRODUCTION OF A COMPACTNESS RATIO

The classification can be made two ways. First, the different classes are based on thresholds of values of parameters as in the manual analysis of Muselli (Muselli, 2000). But these intervals are arbitrarily chosen and we do not really know if they are representative of the best breakdown of the data set. The second way is that there is no interval to define class breakdown but that a class is composed by objects that show similar characteristics. The notion of compactness replaces the thresholds. The more the class is compact the more the objects that composed it are similar. The optimum classification corresponds to the breakdown of the data set in the most compact classes.

The mean compactness ratio of the classes can be expressed as the Global Balanced Variance within class (GBV):

$$V_{GBV} = \frac{1}{N^2} \sum_{i=1}^{C} Eff_i \cdot V_i$$

$N$ : strength of the set to be classified
$C$ : number of classes
$Eff_i$ : strength of the class $i$
$V_i$ : variance of the class $i$ (Lebart, 1982)

Minimize the GBV corresponds to search for the best breakdown of the data set in the most compact classes.

## APPLICATION TO WARD'S CLASSIFICATION

We first applied the concept of the GBV inside a hierarchic cluster method (Lebart, 1982) as used in the weather generator of Muselli (Muselli, 2000).

The study of the GBV evolution according to the level of the tree structure shows an evident minimum (Fig. 1 curve "Initial"). It is the point where the clustering in compact classes is finished. Beyond this point these compact classes gathered themselves.

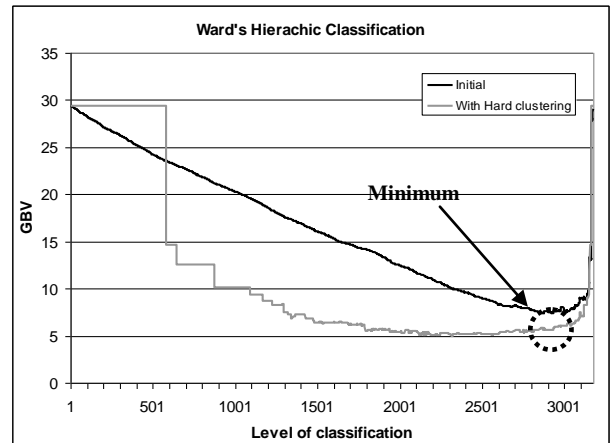

*Figure 1 Compactness ratio for ward's Classification - Station of Le Port (1993 - 2004)*

At this level of the dendrogram, we have N + 1 classes: N compact classes and one class with all the objects that has not been clustered. This last class has a very high variance relatively to the other classes. Studying the characteristics of its own objects, we saw that they do not correspond to extreme values of the original parameters (before PCA). The majority of these objects have intermediary places, between several compact classes.

In order to consider these intermediary objects inside the compact breakdown and optimize the GBV, we complete the hierarchic classification method with hard clustering around defined class centers. This relocation method is considered a classification and it is rarely associated with other ones. But in our case, this association permits the full allocation of all the objects of the data set in clusters and decreases the compactness ratio (Fig. 1 curve "With Hard clustering").

## APPLICATION TO DIDAY'S CLASSIFICATION METHOD

The cross partitioning method (Lebart, 2000) consists in determining the steadiest clusters that compound a data set. The same non hierarchical method is successively applied to the data set. At each run the values of initial parameters are randomized. The different breakdowns, one for each iteration, are compared to evaluate what objects keep the same cluster independently of the initial parameters.

In his method, Diday applied the cross partitioning to successive classifications by Fuzzy C-means (Lebart, 1982). The breakdown of this relocation method depends totally on the initial parameters. The number and the characteristics of the clusters result from the number and the choice of initial the centers.

One time the number of center fixed, the coordinates of the center become the initial parameters. We choose k fuzzy centers (C1, C2 … Ck) and we runs N times the clustering with randomize the initial coordinates of the fuzzy centers. It results N breakdown of the data sets.

We cross the N breakdown to determinate the stable clusters. In every line we affect a statistical combination corresponding to the series of the number of the clusters. The different statistical combinations represent the "stable" breakdowns.

$$\begin{pmatrix} N° & 1st & 2nd & & Nth \\ \text{object} & \text{breakdown} & \text{breakdown} & & \text{breakdown} \\ 1 & 1 & 5 & \cdots & 1 \\ 2 & 1 & 5 & \cdots & 1 \\ \vdots & \vdots & \vdots & & \vdots \\ n & 2 & 1 & \cdots & 5 \end{pmatrix} \Rightarrow \begin{pmatrix} N° & \text{statistical} & \\ \text{cluster} & \text{combination} & \text{strength} \\ 1 & 11\cdots 1 & 15 \\ 2 & 11\cdots 2 & 121 \\ \vdots & \vdots & \vdots \\ i & kk\cdots k & y \end{pmatrix}$$

Even if the cross partitioning method decreases the number of initial parameters, two variables must be observed, the number of fuzzy centers and how many iterations is necessary to get the most stable clustering. We considered two different ways to randomize the coordinates of the fuzzy centers. The first is the generation of $k \times n$ random numbers inside the space of the n dimensions of the k centers. The other is a random selection of k objects in the data set for each runs. Fig. 2 shows the evolution of the compactness ratio for these two methods considering 10 years of data for "Le Port". GBV is closed for the two methods independently of the number of fuzzy centers.

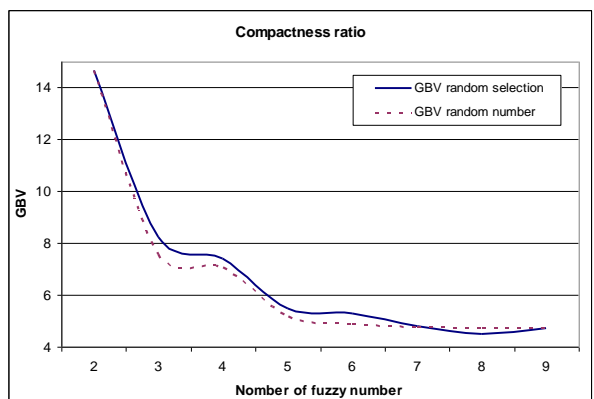

*Figure 2 Compactness ratio for Diday's classification method - Station of Le Port (1993-2004)*

In order to know the best of these methods of fuzzy center choices, an index of stability is introduced. It corresponds to the percentage of objects that are in the same classes from two following steps where we encountered a minimum GBV. Fig. 3 shows the better stability of the process of selection of the fuzzy centers inside the data bases of the classified objects. Around 5 and 6 centers the random selection has its maximum stability. For the random generation of the fuzzy centers coordinates, the stability decreases quickly.

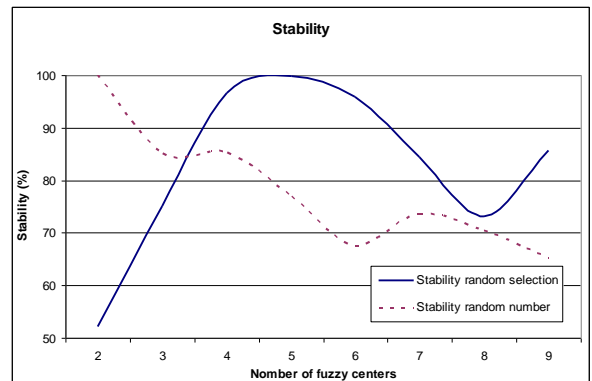

*Figure 3 Stability of Diday's classification method - Station of Le Port (1993-2004)*

For 6 centers and from about twenty iterations, the minimum of the compactness ratio is reached (Fig. 4) whatever is the method of choice of the centers. This minimum corresponds to a high stability (96 %) for the random selection. For further works using the Diday's method, a minimum of at least 20 iterations will thus be necessary.

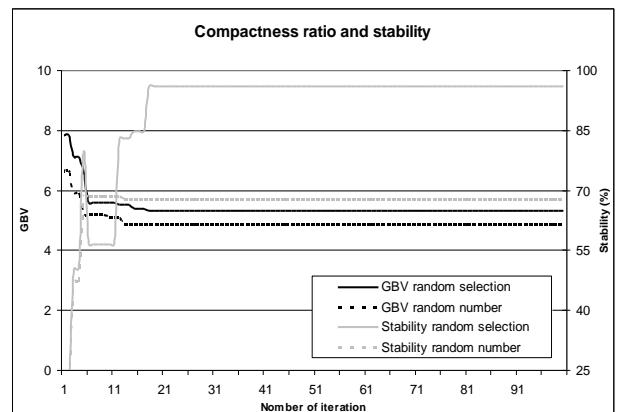

*Figure 4 Compactness ratio and stability of Diday's classification method, 6 fuzzy centers - Station of LePort (1993-2004)*

## COMPARISON

Formally the major difference between these two ways of classification is the approach of the data set. The hierarchic methods show a static and arbitrary breakdown in the cloud of points. The cross partitioning applied to the fuzzy C-means is a dynamic point of view of whole objects. For this comparison, we use the Ward's classification completed by a hard clustering and the Diday's method with 6 fuzzy centers and a maximum of 50 iterations.

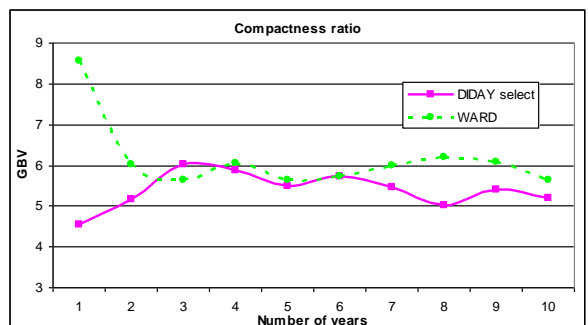

*Figure 5 Compactness ratio comparison of Ward's and Diday's classification - Station of Le Port*

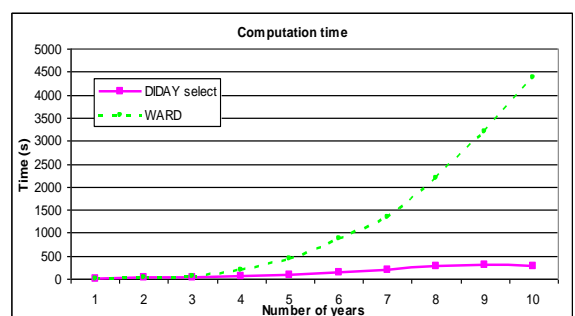

*Figure 6 Computation time comparison of Ward's and Diday's classification - Station of Le Port*

For more than 5 years of data Diday's classification has a lower GBV than Ward's method (Fig. 5). We can consider that the cross partitioning algorithm produces a more compact breakdown than the modified ascendant classification. Computation time is also greatly improved (Fig. 6). The strength of the data set has a lower influence on the Diday's method because the size of the calculation matrix grows in only one dimension. For this non hierarchical classification the computation time depends principally in the initial number of fuzzy centers. For the hierarchic clustering, the two dimensions of the calculation matrix grow as the same time as the strength of the data set.

## ANALYSIS OF THE CLASSIFICATION RESULTS

Regarding previous simulations we retain the Diday's method with 6 fuzzy centers and a maximum of 50 iterations. To test the classification, we use 11 years of data from meteorological station of Le Port situated in the west coast of the island. 5 parameters are considered: air temperature, global radiation, rainfall, wind speed and direction (Table1). For the period 1993 – 2004, a total of 88% of the measured days can be used for this station.

Table 2 shows the correlation between the classes and the global climatic conditions of the Indian Ocean (Robert, 1986). We notice that the 22 classes divide in three groups. There are days of damp season, intermediary days and days of dry season. The intermediary days correspond to the weather we can encounter during dry or damp season and also during the months of transition (April and November). But none of these intermediary days can correspond to dry and humid season in the same time.

After analyzing the frequencies of sequences (right in the table) we find suitable proportions of weather patterns, more than 50 % of time a regime of trade winds and more than 30 % of time a regime of weak pressure gradient.

Some specific synoptic conditions are compounded by different typical days. For example the anti cyclonic regime of trade winds includes 3 classes of intermediary days. Class N°6 corresponds to a regime of trade winds of wet season with a strong period of sunshine (> 5500 W/m in the daytime), temperatures raised (> 28°C) and strong winds in the day directed northeast. Classes 15 and 21 are rather similar. They correspond to a regime of trade winds of dry season with a strong period of sunshine (> 5500 W/m in the daytime), they differ by the orientation of winds and the average temperature. Class 15 has an average temperature of 23.5°C for a southeasterly wind forcing significantly the day. Class 21 has an average temperature of 25°C with a southeasterly wind forcing slightly the day. These last two rather close classes show two directions of trade winds current, east with not enough influence for the considered zone (class 21) and of southeast with a contribution of fresh air (class 15).

*Table 1 Studied locations*

|  |  | **Gillot** | **Le Port** |
|---|---|---|---|
| **Height (m)** | | 22 | 11 |
| **Situation** | | Wind coast | Under the wind coast |
| **Records** | Temperature | x | x |
| | Irradiance | x | x |
| | Hygrometry | x | |
| | Wind | x | x |

*Table 2 Analysis of the correspondences between classes and global synoptic conditions over Indian Ocean.*

| Type of conditions | | Description of the synoptic conditions | Influence on littoral under the wind | Le Port | | | |
|---|---|---|---|---|---|---|---|
| | | | | Humid season | Dry season | Intermediary | Frequencies (%) |
| A | A | High pressure with trade winds. Convection forced along relief giving rain. Included cloudy coat between 1000m and 2500m | Protected by the relief, zone of sun. Trade winds of sector the South or the North for the extremities of the zone | 2, (4) | 20, 17 | 6, 15, 21 | 34.9 |
| B | B | Tropical disturbance of the North in wet season which can give a cyclone. It concerns directly the island of Reunion. Warm and wet air | Cloud layer, following the local nearness of the depression, rains can be strong or diluvian | 8, 14 | | | 0.5 |
| | A perturbed B | The disturbance forms in the part South West of the Indian Ocean and do not concern directly Reunion. The direction of trade winds remains almost unchanged | Protected by the relief, zone of sun. Trade winds of sector the South or the North for the extremities of the zone | 3, (4) | | 11, 22 | 8.6 |
| | Off season disturbance | Depression forming in the North of Reunion in the low coats of the atmosphere | Strong rains and wind of sector the South to the East | | 16 | | 0.6 |
| C | C | Cold front of the South created by a polar depression which concerns directly the island of Reunion. Fresh and wet air | Strong rains especially in the South with presence of wind and strong swell from the South | | 19 | | 1.2 |
| | A perturbed C | Cold front of the South or the South West in approach of Reunion. Decline of the regime of trade winds | Weak rains to strong especially in the part the North | | | 18 | 6.9 |
| D | D1 | Undulation of trade winds and thalweg of height (running from the North in height) provoke a North/South directed cloudy bar moving in the course of trade winds East/West | Protected by the relief, zone of sun | | | 5, 13 | 26.5 |
| | D2 | Axis of instability directed North/South due to the contact of 2 air masses, generally a high and a low pressure mass, it moves of the East to the West | Rains except for the zone the North West | | | 5, 13 | |
| E | E | Weak pressure gradient leaving place with a free convection. Regime of breeze with forming of rainy clouds on relief | No rain and regime of thermal breezes | 1, 7 | | 5, 13 | |
| | E perturbed A | Weak pressure gradient with a light stream of trade winds | No rain and regime of thermal breezes | | | 5, 13 | |
| | E perturbed B | Further to a disturbance of the regime of trade winds ( A ) by a tropical depression ( B ) settles down a regime of weak pressure gradient. Warm and wet air | Possibility of thunderstorms, regime of thermal breezes | 9 | | 12 | 9.8 |
| | E perturbed C | Further to a disturbance of the regime of trade winds ( A ) by a polar depression ( C ) settles down a regime of weak pressure gradient. Fresh and wet air | Good weather in the north and cloudy in the South, the weak rainy risk, no wind | | 10 | | 10.9 |

## GENERATION OF A TYPICAL WEATHER YEAR

In order to test the quality of the classification process, an artificial typical weather year is generated from the classes of weather sequences.

In this study we focus on the capacity of the selected classification to produce the best entries for the generation process. We approach only briefly the method of generation of the artificial year.

First order Markov Chain Model (MCM) has already been used and successfully applied to the reproduction of daily weather chains. Associated with a specific artificial neural network (ANN or simply NN) in order to obtain time steps shorter than a day, a methodology was developed inside Runeole to generate typical weather files (David, 2005). The complete analysis of climatic data using Principal Component Analysis (PCA), classifications and basics statistics has supplied a large amount of entries for MCM and NN.

## APPLICATION IN THE NUMERIC SIMULATION OF BUILDINGS

To lead the validation of this method, we need to test the capacity of the method to generate 1 typical year with the same characteristics as a long period. A multi-site approach is necessary to evaluate the method under different climates.

To compare the different databases used for the sizing and the long-term energy forecasts, we make a comparison of simulations of 11 years (1993-2004) of measured data and typical year generated from the results of the classification. We use the weather stations of Gillot, altitude 22 meters, where more than 68.19% of the measured days are available and La Plaine des Cafres, altitude 1560 meters, where 77.32% of the measured days are available. These two places experience very opposite microclimates, hot and windy for Gillot and cold and rainy for La Plaine des Cafres. Table 1 shows the available hourly records for these two places. The weather station of Le Port no includes the hygrometry measurement and we will not take it into account for this place.

*Table 3 CODYRUN inputs and outputs*

| Inputs | | Outputs |
|---|---|---|
| Weather file | Building description | Result file |
| Air temperature (°C) | The building description is elaborated inside the CODYRUN user interface : | Air temperature (°C) |
| Beam radiation (W/m²) | | Walls temperature (°C) |
| Diffuse radiation (W/m²) | | Radiation through windows (W/m²) |
| Relative humidity (%) | Thermal zones | Specific humidity (kg) |
| Solar zenith angle (°) | | Air mass exchange (kg/s) |
| Solar azimuth (°) | Walls, roofs, windows, floors, … | Cooling loads (Wh) |
| Wind speed (m/s) | Internal loads | Heating loads (Wh) |
| Wind direction (°) | Air conditioning | |

An experimental single zone already used for thermal issues (Mara, 2001) (Miranville, 2003) under tropical conditions is studied (Fig. 7) with the thermal simulator CODYRUN (Boyer, 1995). CODYRUN is a multiple-zone building thermal simulation software regrouping design and research aspect. From building description and weather data it provides results that characterized the indoor thermal ambiance and the energy consumption of the air conditioning (Table 3). The test cell is a cubic-shaped building with a single window on the south wall and a door on the north one. It would be interesting to complete the validation with simulations of different types of buildings. For the following comparison, we only use a test cell with a weak thermal inertia.

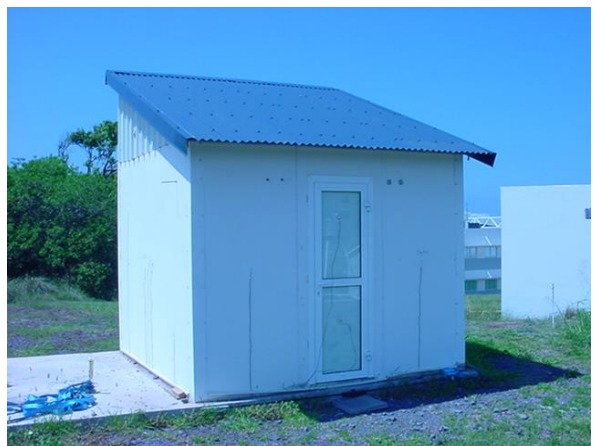

*Figure 7 "LGI" cell*

In Tables 4 and 5 are indicated the errors for the main statistical indicators between 11 measured years and the year generated artificially. For this comparison we use weather data with an hourly time step. We notice that the process will give the average values for all the climatic parameters with an error lower than 1 %. On the other hand, extremes are not so well restored. But they are rather close to the main variables, the temperature, the irradiance and the humidity. The strong distances on certain extremes result from the rarity of their frequency. For example, a wind of more of 50 m/s does not reproduce only once every 10 years, that explains the 69 % error for the site of Gillot (Table 4) between the artificial and real maximum data for the hourly wind speed. Other strong errors on extremes also result from a bad filtering of the measured data base. For example, the minimum of relative humidity measured for site of Gillot (2%) is hardly. It results in an error of 94% (Table 4) with the artificial data. The 100 % error is a zero value not reached by the artificial data. The speed and the direction of wind represent it well.

*Table 4 Weather files comparison between 11 measured years and 1 artificial typical year - Station of Gillot*

| Error (%) | Temperature | Irradiance | Wind speed | Wind direction | Relative Humidity |
|---|---|---|---|---|---|
| Annual mean | 0.23 | 0.68 | 0.5 | 0.84 | 0.23 |
| Maximum | 7.58 | 10.77 | 68.86 | 44.44 | 7 |
| Minimum | 21.08 | 0 | 100 | 100 | 94.23 |
| Standard deviation | 30.64 | 5.5 | 51.37 | 58.01 | 33.23 |

*Table 5 Weather comparison between 11 measured years and 1 artificial typical year - Station of La Plaine des Cafres*

| Error (%) | Temperature | Irradiance | Wind speed | Wind direction |
|---|---|---|---|---|
| Annual mean | 0.86 | 1.12 | 9.89 | 2.20 |
| Maximum | 14.01 | 5.69 | 75.31 | 16.67 |
| Minimum | | 0 | 100 | 100 |
| Standard deviation | 8.64 | 3.52 | 23.68 | 28.70 |

The simulation of 10 years of measured data and the artificial typical year with CODYRUN allows us to make even comparisons as with the weather files. The Tables 6 and 7 show the relative errors for the various parameters characterizing the atmosphere inside the test cell under the two microclimates of Gillot and La Plaine des Cafres. We ran first simulations without using air treatment in order to compare the passive behavior of the building. The second simulations were carried out with a constant internal ambiance to evaluate the energy loads of air conditioning systems for these two places.

We notice a very good correlation for the average and the extremes of the temperature and the humidity. But the strong distances on the minimum of humidity result from a weakness of the filtering!

The energy consumptions of air conditioning were simulated for a constant temperature of 20°C and 70% humidity. The error between long term and the artificial year (Tables 6 and 7) remains weak for the mean consumptions and maximum is widely underestimated.

*Table 6 Simulation comparison between 11 measured years and 1 artificial typical year - Station of Gillot*

| Error (%) | Temperature | Specific Humidity | Cooling load | Heating load |
|---|---|---|---|---|
| Annual mean | 2.31 | 0.73 | 10.11 | 61.14 |
| Maximum | 12.81 | 7.32 | 34.78 | 74.14 |
| Minimum | 11.74 | 92.14 | 0 | 0 |
| Standard deviation | 12.70 | 16.03 | 20.85 | 68.32 |
| Percentage of system on time | | | 90 | 10 |

*Table 7 Simulation comparison between 11 measured years and 1 artificial typical year - Station of La Plaine des Cafres*

| Error (%) | Temperature | Cooling load | Heating load |
|---|---|---|---|
| Annual mean | 2.96 | 33.19 | 3.43 |
| Maximum | 21.19 | 47.40 | 22.29 |
| Minimum | 26.17 | 0 | 0 |
| Standard deviation | 13.15 | 41.56 | 5.35 |
| Percentage of system on time | | 25 | 75 |

The simulation results show that the typical artificial year reproduces a long term period without significant errors on averages. Even if the extreme values of the climate are underestimated, the long term behaviour of building is well predicted thanks to the typical artificial weather year.

## CONCLUSION

The automation of the methods of classification as well as their optimization offers new perspectives. The strength of data to be feigned to take into account the extreme and the means characters of weather can be significantly reduces by the generation of a typical year from determined classes. The method of cross partitioning allows us to treat a very large range of data with realistic computation times.

The classification also gives us information about correlations between the parameters in well determined conditions. The number of classes can be one of the entries for a model of simultaneous generation of climatic data. It opens the possibility of generated various variables simultaneously.

A deepening of the method of generation is propose in a follow-up study (David,2005) about the integration of the number of class in the entry of a generator multi variable by a specific Neural Network and Markov Chains (Muselli, 2001).

Runeole has been used for the determination of specific climatic areas in Reunion Island and for the elaboration of the future thermal standard in the French overseas territories.

Future work will be devoted to the spatial interpolation of the artificial typical years. In Reunion Island, the large amount of microclimates will permit to test if it is possible to use model from the literature. The method developed by Wilks (Wilks, 1999) on the adaptability of the random process of Markov chains seems to be applicable.

## ACKNOWLEDGMENT

This research is supported by a grant from the Conseil Regional of Reunion Island. The weather data was supplied by Meteo France French weather services in Reunion Island.